\documentclass[prr,aps,amssymb,twocolumn,superscriptaddress,showkeys,notitlepage,nobalancelastpage]{revtex4-1}
\usepackage{graphicx}
\usepackage{dcolumn}
\usepackage{bm}
\usepackage{ulem}
\usepackage{mhchem}
\usepackage{changepage}
\usepackage{float}
\usepackage{amsmath}
\usepackage{xcolor}
\usepackage{bbold}
\usepackage{mathtools}
\usepackage{multirow}
\usepackage{comment}

\usepackage{hyperref}

\newcommand{\kvec}{\mathbf{k}}
\DeclareUnicodeCharacter{03B2}{$\beta$}

\begin{document}
\begin{abstract}

FeSe has been one of the most intensively studied iron-based superconductors over the past two decades, exhibiting a wide range of phenomena such as unconventional superconductivity, nematic order, magnetism, orbital-selective correlations, and structural phase transitions. While topologically non-trivial phases have been identified in certain cases—such as Te-doped FeSe and monolayer FeSe—topology in bulk FeSe has largely remained unexplored.
In this work, we propose a new route to realize topological phases directly in bulk FeSe. We demonstrate that breaking the tetragonal $C_4$ rotational symmetry, thereby lowering the crystal symmetry, can drive FeSe into a strong topological insulating phase. To support this, we perform density functional theory calculations and analyze the band structure using Topological Quantum Chemistry and symmetry-based indicators.
Our results show that both uniaxial strain and temperature-induced structural changes lead to non-trivial band topology. Moreover, incorporating electronic correlations through dynamical mean field theory  reveals that the topological characteristics near the Fermi level remain robust, as the relevant bands experience only moderate renormalization.
These findings highlight strain as a promising mechanism to induce topological phases in FeSe.

\end{abstract}

\title{Symmetry-breaking-induced topology in FeSe}

\author{Mikel García-Díez}
\thanks{These authors contributed equally to this work.}
\affiliation{Donostia International Physics Center, Paseo Manuel de Lardizabal 4, 20018 Donostia-San Sebastian, Spain.}
\affiliation{Physics Department, University of the Basque Country (UPV/EHU), Bilbao, Spain}

\author{Jonas B.~Profe}
\thanks{These authors contributed equally to this work.}
\affiliation{Institut f\"ur Theoretische Physik, Goethe-Universit\"at Frankfurt, 60438 Frankfurt am Main, Germany}

\author{Augustin Davignon}
\affiliation{D\'epartement de Physique et Institut Quantique, Universit\'e de Sherbrooke, Sherbrooke, J1K 2R1, Qu\'ebec, Canada.}

\author{Steffen Backes}
\affiliation{RIKEN iTHEMS,  Wako, Saitama 351-0198, Japan; Center for Emergent Matter Science, RIKEN, Wako, Saitama 351-0198, Japan}

\author{Roser Valent\'i}
\email{valenti@itp.uni-frankfurt.de}
\affiliation{Institut f\"ur Theoretische Physik, Goethe-Universit\"at Frankfurt, 60438 Frankfurt am Main, Germany}

\author{Maia G.~Vergniory}
\email{maia.vergniory@usherbrooke.ca}
\affiliation{Donostia International Physics Center, Paseo Manuel de Lardizabal 4, 20018 Donostia-San Sebastian, Spain.}
\affiliation{D\'epartement de Physique et Institut Quantique, Universit\'e de Sherbrooke, Sherbrooke, J1K 2R1, Qu\'ebec, Canada.}
\affiliation{Regroupement Qu\'eb\'ecois sur les Mat\'eriaux de Pointe (RQMP), Quebec H3T 3J7, Canada}

\date{\today}

\maketitle

{\it Introduction.-} Iron-based superconductors (FeSCs) are among the most studied families of superconducting materials since their discovery in 2008~\cite{takahashi2008, fernandes2022iron}. Within this family, \ce{FeSe} has attracted particular attention~\cite{hsu2008,glasbrenner2015, bohmer_variation_2016, coldea2018key, benfatto_nematic_2018, kang_superconductivity_2018, bohmer_nematicity_2018, kreisel2020remarkable, baek_orbital-driven_2015, bartlett_relationship_2021} due to its unique properties: it exhibits unconventional superconductivity~\cite{sprau2017discovery,kreisel2020remarkable}, nematic order~\cite{bohmer2016electronic,chubukov2015origin,unidirectional-nematic-bond}, strong orbital-selective correlations~\cite{yin2011kinetic,effect-correlations-expansion, hubbard-like-bands,yu2018orbital},  and 
lacks long-range magnetic ordering at ambient pressure~\cite{magnetic-gs}.
Moreover, \ce{FeSe} lies at the boundary of multiple competing phases, such that modest external perturbations can efficiently induce charge order~\cite{Chen2023}, magnetic order~\cite{pressure-induced-afm}, and significantly modulate its superconducting behavior~\cite{pressure-critical-field, suppression-orbital-chemical, maximizing-tc, evolution-hightc}.

The remarkable tunability of FeSe stems from its electronic band structure, which is highly sensitive to minor changes in the crystal structure, such as variations in the Fe–Fe and Fe–Se bond lengths~\cite{glasbrenner2015,basic-properties}. This structural sensitivity allows for the modulation of FeSe's electronic properties with relatively small external stimuli, making it more responsive compared to other compounds~\cite{strain-tuning,pressure-critical-field,suppression-orbital-chemical,maximizing-tc,evolution-hightc,basic-properties,pressure-induced-afm,electronic-magnetic-phase-diagram,crystal-orientation,PhysRevB.91.041112}. Specifically, these interatomic distances can be tuned by applying pressure or strain. For instance, lattice expansion has been theoretically shown to trigger a Lifshitz transition, boost local magnetic moments, and intensify orbital-selective correlations, leading to a stronger quasiparticle mass renormalization~\cite{lifshitz-tetragonal}. On the other hand, compressive pressure, beyond 1.6 GPa, suppresses nematic fluctuations~\cite{scherer2017interplay,collapse-nematic,PhysRevLett.123.167002}, reduces the quasiparticle mass enhancement~\cite{Skornyakov_2018}, and drives the system towards an orthorhombic phase with stripe-like magnetic ordering.

In the search for topological signatures in \ce{FeSe} and its doped siblings, various proposals have been put forward by making use of the above-mentioned extreme tunability.
Chemical doping, in particular, has proven effective in driving topological transitions. For example, substituting selenium with iodine and tellurium in
\ce{FeSe_{0.325}I_{0.175}Te_{0.5}} has been shown to produce a Dirac cone at the Fermi level~\cite{band-engineering}. Similarly, 
 \ce{FeSe_{0.45}Te_{0.55}}
undergoes a band inversion, resulting in a $\mathbb{Z}_2$ topological insulator (TI) phase characterized by a surface Dirac cone~\cite{topological-nature,orbital-selective-topology}. In \ce{FeSe_{0.45}Te_{0.55}}~\cite{Zhang20182, Wang2018, Wang2020, topo-surfcon-fts, Li2021, Machida2019} surface magnetism and time-reversal symmetry (TRS) breaking have been proposed to enable the emergence of Majorana zero modes and a surface Dirac cone. Apart from this, single-layer FeSe, which exhibits a strikingly high $T_c$ of $\sim$65~K up to $\sim$100~K~\cite{topo-single-layer,Feng2018, edge-states-film}, has
been suggested to host high-order topological states with antiferromagnetic (AFM) magnetization~\cite{fragile-flat-band}. Furthermore, when placed on a SrTiO$_3$ substrate, it can also host one-dimensional robust topological edge states protected against any TRS-preserving perturbation within 40 meV of the Fermi level~\cite{edge-states-film,Yuan2018} reportedly due to strain induced by the substrate. Additionally, higher-order topological phases have also been reported in this configuration by
forcing a canted AFM state~\cite{higher-order-monolayer}. Theoretically, it has also been proposed that some of the Fe-based superconductors have a correlation-driven topological phase~\cite{Corrtopology, PhysRevB.109.L241106}. These proposals render the \ce{FeSe} family as a promising contender for the unambiguous identification of Majorana zero modes and the construction of topological quantum computers~\cite{RevModPhys.80.1083}.

\begin{figure}
    \centering
    \includegraphics[width=0.9\linewidth]{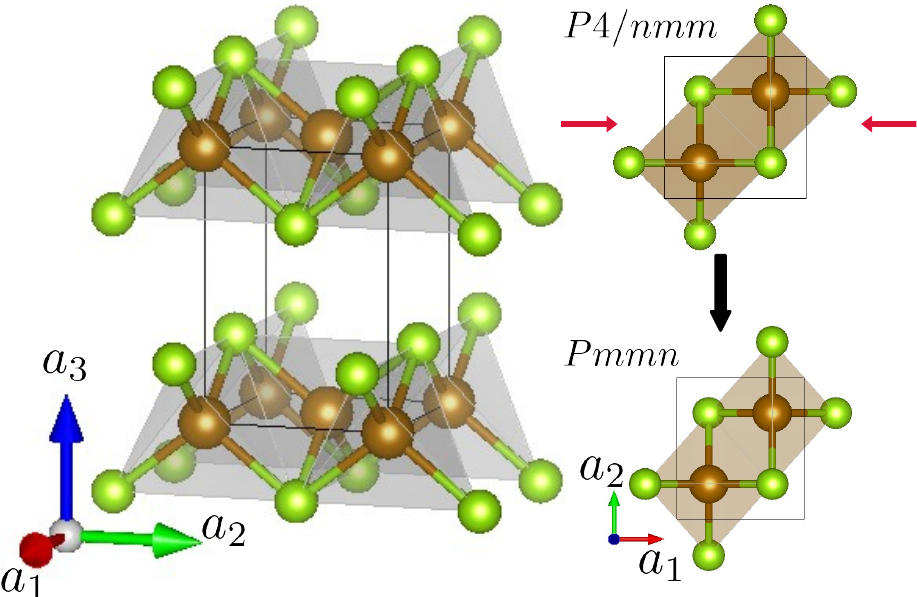}
    \caption{Atomic structure of FeSe. Strain is applied along the ${a_1}$ direction of the unit cell (red arrows) of the uncompressed tetragonal structure belonging to space-group (SG) No.~129. Once uniaxial strain is applied, the systems adopts an orthorhombic symmetry described by SG No.~59 (or 67). The length of the ${a_2}$ axis is exaggerated to better display the orthorhombic symmetry. 
    The parameters of the tetragonal and orthorhombic phases are specified in the appendix or supl. inf.}
    \label{fig:structure}
\end{figure}

In view of these developments, in this work we propose another path to engineer topological phases in FeSe, namely breaking the tetragonal $C_4$ rotational symmetry by e.g.~in-plane uniaxial strain~\cite{uniaxial-strain}. 
To analyze the topology, we perform electronic structure calculations
of \ce{FeSe} in both its tetragonal (strained and unstrained) and orthorhombic phase. Starting from density functional theory (DFT), we classify the topology of the resulting bands using topological quantum chemistry (TQC)~\cite{tqc,mtqc} and symmetry indicators~\cite{mtqc,sis-song} (SI). We predict that uniaxial compression (or expansion) in the tetragonal structure along the $a_1$ or $a_2$ axis (see Fig.~\ref{fig:structure}) as well as the structural transition to the orthorhombic phase upon lowering temperature enable a topological phase transition into a strong TI. The mechanism behind this topological transition is the reduction in symmetry, leading to different elementary band representations (EBRs). To verify that the strong correlations in \ce{FeSe} do not invalidate these predictions, we subsequently perform charge-self-consistent dynamical mean-field theory (DMFT) calculations. Here, we observe that the bands near the Fermi level are still identifiable with their DFT counterparts and that no new crossings appear between them. These findings indicate that the predicted topological states are not destroyed by correlations, establishing \ce{FeSe} as a platform to study topology in a correlated metal. 

{\it Methods.-} Electronic structure calculations were performed within DFT with the Vienna \textit{Ab-initio} Simulation Package (VASP) software~\cite{vasp1,vasp2}, choosing Projector Augmented Wave (PAW) pseudo-potentials~\cite{paw} with the revised Perdew-Burke-Ernzenhof implementation of the Generalized Gradient Approximation (GGA) for the exchange-correlation functional for solids (PBESOL)~\cite{gga_pbe, PhysRevLett.100.136406}. We performed all calculations in the fully relativistic setting, including spin-orbit coupling. Furthermore, we include van der Waals corrections in the form of DFT-D3 with Becke-Johnson damping~\cite{Grimme2011}. The energy cutoff of the plane wave basis is set to 800 eV. As a $\kvec$ point grid we used a regular Monkhorst-Pack point set centered at $\Gamma$ of dimensions $11\times 11 \times 10$. Throughout the calculations we used a gaussian broadening of $0.05$ eV. Structural optimization of the ion positions was performed using the Residual Minimization Method with Direct Inversion in the Iterative Subspace (RMM-DIIS) implemented in VASP until all forces exerted on the atoms had a modulus smaller than $10^{-3}$ eV/\AA. Strained structures are relaxed using the lattice constraints feature of VASP. Band structure plots were prepared using VASPKIT~\cite{Wang2021}. All simulation data is provided in a NOMAD dataset~\cite{NOMAD_dataset}.

\begin{figure*}[!bht]
    \centering
    \includegraphics[width=\linewidth]{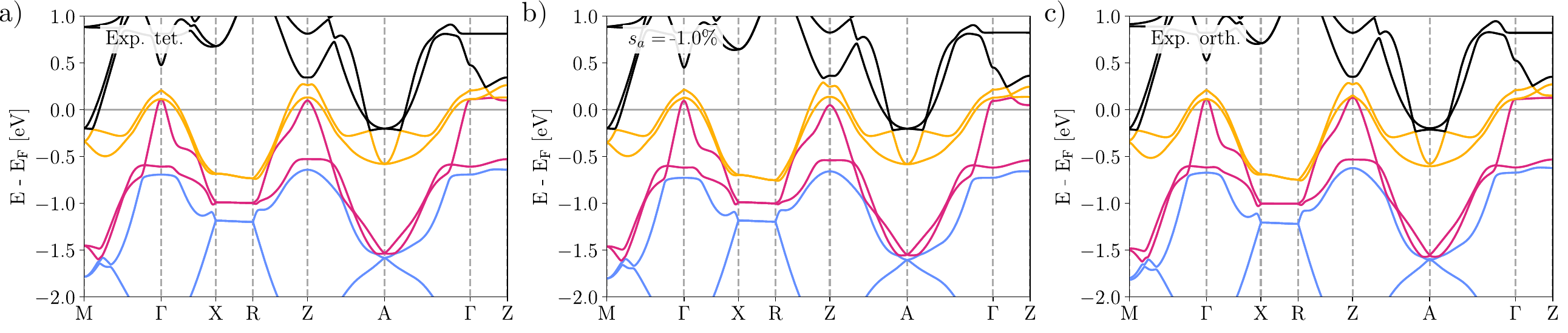}
    \caption{Electronic band structure of \ce{FeSe}  under a) ambient conditions in the high temperature tetragonal phase, b) compressive strain along the $a_1$ axis  and c) under ambient conditions in the low temperature orthorhombic phase. The band coloring are a guide to the eye and indicate fillings of 20 (blue), 24 (magenta) and 28 (yellow) electrons respectively. Under strain along $a_1 = a$ or under strain along $a_1a_2 = ab$ the space group changes to an orthorhombic phase (SG No.~59 and SG No.~67 respectively) . We retain the labeling of the irreducible path of the unstrained case for both strained cases easier comparability. }
    \label{fig:bandstructures}
\end{figure*}

To identify the symmetry of the electronic eigenstates, we used the IrRep package~\cite{irrep}. Consequently, all the labeling conventions agree with the information from the Bilbao Crystallographic Server~\cite{double_space_group}.
In order to construct a basis set of maximally localized Wannier functions for any of the crystal structures, we employed Wannier90~\cite{wannier90} projecting onto the $d$ orbitals for Fe and $p$ for Se. This results, after the wannierization process, in a total of 32 basis functions, taking into account the doubling due to spin. A frozen disentanglement window of approximately 2 eV centered at the corresponding Fermi level was chosen to reproduce the low energy spectrum and the gaps where surface states appear. Further calculations using the interpolated tight-binding model in the Wannier basis were performed in WannierTools~\cite{wanniertools}. Surface states were computed in a semi-infinite slab geometry via the iterative Green's function method with 3 principal layers, as implemented also in WannierTools, obtaining the surface and bulk density of states (DOS).

Since \ce{FeSe} is a strongly correlated metal~\cite{hubbard-like-bands,PhysRevB.82.064504}, DFT within the framework of conventional exchange-correlation functionals neither predicts the correct crystal structure in comparison with the experimental structure~\cite{PhysRevB.89.035150, PhysRevLett.115.106402, Skornyakov_2018}  nor provides an accurate description of the electronic properties as observed in e.g. ARPES measurements~\cite{hubbard-like-bands,PhysRevB.82.064504}, even when taking as an input the experimental structure~\cite{nonloc_4}.  In order to overcome these shortcomings, we performed 
charge self-consistent dynamical mean-field theory (cscDMFT) calculations to account for correlation effects in the electronic properties of \ce{FeSe} beyond DFT. With this, we can assess whether the diagnosed topology within DFT is destroyed by electronic correlations or not. Furthermore, to obtain realistic crystal structures under strain, we extracted the strain tensor from DFT and applied it to the experimentally-measured structure~\cite{topological-nature,Koz2014}. For more details on this procedure, see Appendix~\ref{App::relax}.

For the DFT+DMFT calculations we first converged a DFT calculation using Wien2k v23.1~\cite{Wien2k_a,Wien2k_b} employing a GGA~\cite{gga_pbe} functional. Since the main goal of these calculations is to get a general grasp of the renormalization effects due to correlations beyond GGA, we do neglect spin-orbit coupling as for 3$d$ electrons has a small contribution and would substantially complicate the DMFT computations. These calculations were performed for the same set of crystal structures analyzed with VASP and we ensured that at the DFT level both Wien2k and VASP provide the same results. We used $(\vec{R}\vec{k})_{\rm{max}} =8$ and a Monkhorst-Pack point set of 5000 $\mathbf k$-points. The DMFT part was performed with the eDMFT code~\cite{PhysRevB.81.195107} in combination with a continuous-time quantum Monte Carlo~\cite{PhysRevB.75.155113} solver for the impurity problem. We used the full Slater parametrization of the interaction with $U=4.6\,$eV and $J=0.7\,$eV~\cite{Miyake2010}. DMFT calculations were performed at a temperature of $T=150$~K. We used fixed projections onto real harmonics such that the DFT+DMFT algorithm becomes stationary~\cite{station_DMFT} and utilized the exact double counting scheme~\cite{PhysRevLett.115.196403}. To calculate the spectral function, we perform an analytic continuation both with the maximal entropy algorithm~\cite{PhysRevB.96.155128} and performing a Fermi-liquid fit of the self-energy. We ensured that both methods agree near the Fermi-level.

{\it Electronic structure at ambient pressure and high temperature.-} Under ambient conditions and at temperatures above 90K, \ce{FeSe} crystallizes in a tetragonal structure described by the space group (SG) $P4/nmm$ (No.~129 in the Belov–Neronova–Smirnova (BNS) notation~\cite{bns_notation}), see Fig.~\ref{fig:structure}. This space group is generated by the identity, lattice translations, and the symmetry operations $\{4^+_{001}|1/2,0,0\}$, $\{2_{010}|0,1/2,0\}$, and inversion $\{\bar{1}|0,0,0\}$, written in Seitz notation. The conventional unit cell contains two Fe atoms at the $2a$ Wyckoff position (WP) and two Se atoms at the $2c$ WP.

Previous studies~\cite{topological-nature,basic-properties} have shown that the band structure -- particularly along the high-symmetry line $\Lambda$ connecting $\Gamma = (0,0,0)$ and $Z = (0,0,1/2)$ in fractional coordinates (see Fig.~\ref{fig:bandstructures}) -- is highly sensitive to the internal $z$ position of the Se atoms located symmetrically at $z$ and $-z$ within the unit cell.  As a result, isoelectronic doping of \ce{FeSe} with \ce{Te} atoms at the chalcogen sites modifies the electronic dispersion along $\Lambda$. This substitution, as mentioned above, has been demonstrated to induce a topological phase transition via band inversion~\cite{topological-nature}. In essence, the incorporation of \ce{Te} alters the coupling between chalcogen atoms in adjacent unit cells along the ${a}_3$ direction, both by shifting the $z$-coordinate of the \ce{Te}/\ce{Se} atoms and by increasing the interlayer orbital overlap due to the more extended $p_z$ orbitals of \ce{Te}~\cite{topological-nature}.

\begin{figure}
    \centering
    \includegraphics[width=\linewidth]{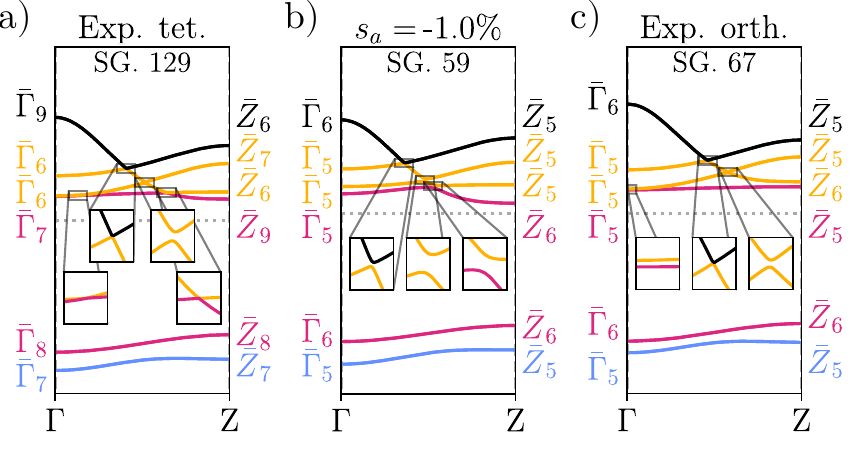}
    \caption{Zoom in on the $\Gamma-Z$ line of the band structures under a) ambient conditions in the high temperature tetragonal phase, b) compressive strain along the $a_1$ axis and, c) under ambient conditions in the low temperature orthorhombic phase. The band coloring are a guide to the eye and indicate fillings of 20 (blue), 24 (magenta) and 28 (yellow) electrons respectively. We indicate the irreducible representations at the two high symmetry points on the left and right axis respectively. In the insets we zoom in on the band crossings and avoided crossings to make visual identification of topological gaps possible. Note that in panel c) the gap in the central inset is $\approx 0.1$meV.}
    \label{fig:zoom}
\end{figure}

In contrast to the Te-doped case,  the structure of \ce{FeSe} at ambient pressure does not exhibit a gap near the Fermi level for a filling of 28 electrons (total number of valence electrons in \ce{FeSe} up to the Fermi level)\footnote{For our choice of pseudopotentials, the filling of 28 electrons corresponds to undoped FeSe.}.
As shown in Fig.~\ref{fig:zoom}~a), the top magenta-colored band and the lowest yellow-colored band at the $Z$ point correspond to the irreducible representations (irreps) $\bar{Z}_9$ and $\bar{Z}_6$, respectively. These irreps subduce to $\bar{\Lambda}_7$ and $\bar{\Lambda}_6$ along the $\Gamma$–$Z$ high-symmetry line and therefore cannot hybridize due to their distinct symmetry properties~\cite{fgt-nodal-lines}. As a consequence, a topological classification near the Fermi level is not well-defined in this regime~\footnote{Throughout this work, we use the term “filling” to refer to the band index, rather than the actual electron count obtained by integrating the density of states (DOS) up to the chemical potential.}.

\begin{table}[]
\begin{tabular}{c|c|c|c|c}
Structure                            & SG                                     & 28 electrons          & SIs                                     & TI                       \\ \hline
\multirow{2}{*}{Tetragonal structure} & \multirow{2}{*}{129}  &  \multirow{2}{*}{ES}  & \multicolumn{1}{c|}{$z_{2w,3}=0$}       & \multirow{2}{*}{NT}  \\
                                     &                      &                        & \multicolumn{1}{c|}{$z_4=0$}            &                          \\
\multirow{2}{*}{$a_1$ compression 1\%} & \multirow{2}{*}{59}  & \multirow{2}{*}{NLC}  & \multicolumn{1}{c|}{$z_{2w,3}=1$}       & \multirow{2}{*}{Strong}  \\
                                     &                      &                        & \multicolumn{1}{c|}{$z_4=1$}            &                          \\
\multirow{2}{*}{Orthorhombic structure} & \multirow{2}{*}{67} & \multirow{2}{*}{NLC} & \multicolumn{1}{c|}{$z_{2w,3}=1$}       & \multirow{2}{*}{Strong} \\
                                     &                      &                       & \multicolumn{1}{c|}{$z_4=1$}            &                                                   
\end{tabular}
\caption{Symmetry and topology for \ce{FeSe} at ambient conditions and under in-plane strain. The first column indicates the considered structure with their corresponding SG in the second column. The third column are the irrep decomposition: Enforced Semimetal (ES), non topological (NT) and non-linear combination of EBRs (NLC). The columns 4-5 SIs and the TI classification for the 28 electron filling.}
\label{tab:phase_space}
\end{table}

\begin{figure*}
    \centering
    \includegraphics[width=\linewidth]{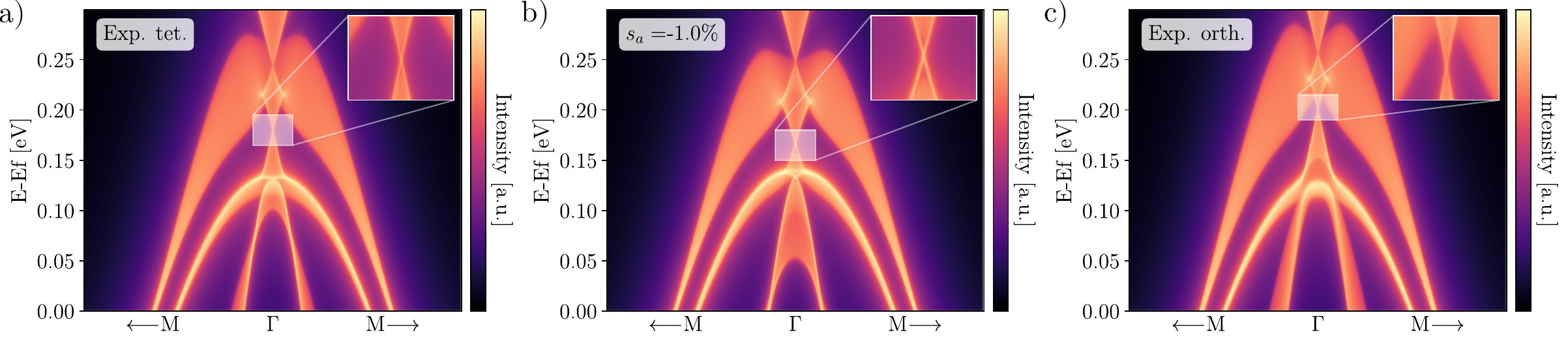} 
    \caption{Momentum resolved density of states for a semi-infinite slab in the $ a_3$ direction in a) the high temperature tetragonal phase, b) compressive strain along the $a_1$ axis and, c) under ambient conditions in the low temperature orthorhombic phase along the path $2/5M\rightarrow \Gamma \rightarrow 2/5M$. The data was computed using the Wannier tight-binding model. The insets show a zoom into the surface Dirac cones present in all three structures due to the strong $\mathbb Z_2$-odd topology.}
    \label{fig:surface-states}
\end{figure*}

{\it Compression along the $a_1$ axis.- } When applying strain along the $a_1$ (or $a_2$) axis, the band structure changes slightly along the $\Gamma-Z$ line (see 
Fig.~\ref{fig:bandstructures}~b) compared to Fig.~\ref{fig:bandstructures}~a)). Additionally, the symmetry is lowered to an orthorhombic phase described by SG $Pmmn$ (No.~59). The crucial difference between the SGs is that the in-plane four-fold rotational symmetry axis, which was present in the tetragonal phase, is lifted. This change of symmetry implies that the two-dimensional little group irreducible representations at $\Gamma$ and $Z$ are subduced to the irreps of the new space group following
\begin{equation}
\label{eq:subduce}
    \bar\Gamma_6,\bar\Gamma_7 \to \bar\Gamma_5 \quad \bar\Gamma_8,\bar\Gamma_9\to \bar\Gamma_6\,.
\end{equation}
As a result of the strain and the subsequent symmetry reduction, the dispersive band --now exhibiting stronger dispersion compared to the unstrained structure--  can hybridize with the flatter bands and gap out. The set of four disconnected bands, see Fig.~\ref{fig:zoom} marked in blue, corresponding to a filling of 16 to 20 electrons constitutes a split EBR~\cite{PhysRevB.97.035139}. The next four bands, see Fig.~\ref{fig:zoom} marked in magenta, correspond to a non-linear combination (NLC) of EBRs. Consequently, the energy gap at 24 electrons is topological, since the collection of the lowest 24 bands cannot be expressed as a sum of EBRs with positive integer coefficients. This implies that there exists no atomic limit consistent with these bands, confirming their non-trivial topology. The next four bands,  see Fig.~\ref{fig:zoom} marked in yellow, again constitute a split EBR. Therefore, the experimental filling, which corresponds to 28 electrons, is a strong topological insulator. Notably, the key component to drive the system to the topological phase is the reduced point group symmetry as already in the tetragonal structure we find a strong TI, when performing the analysis for subgroups of SG.~129.

\begin{figure*}
\includegraphics[width=\linewidth]{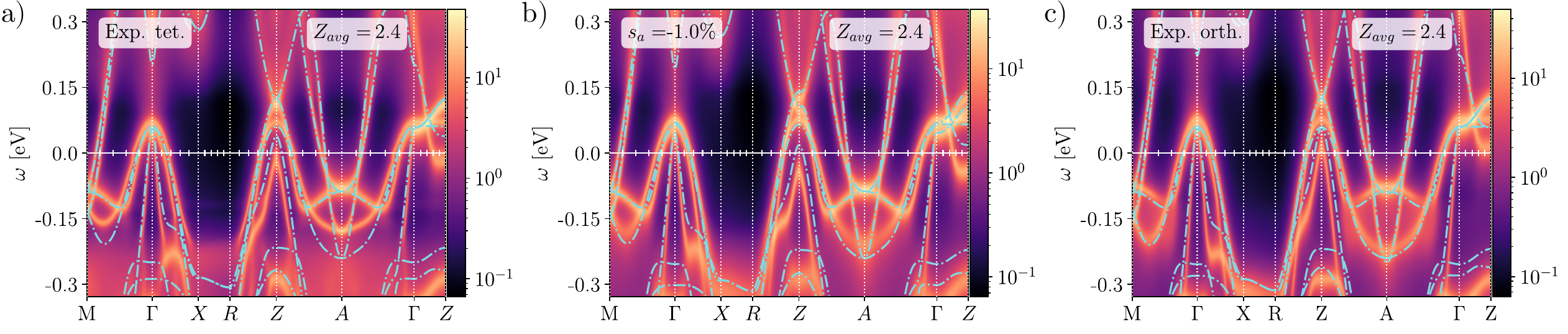}
    \caption{Spectral functions of \ce{FeSe} under ambient conditions in a) the high temperature tetragonal phase, b) compressive strain along the $a_1$ axis, and c) under ambient conditions in the low temperature orthorhombic phase. The charge self-consistent DMFT calculations were performed at $T=150$K. The cyan dashed lines indicate the DFT band structure, renormalized by the average mass-enhancement of the DFT+DMFT calculations. In addition to the renormalization, we observe that the bands mainly belonging to orbitals not included in the impurity problem are shifting with respect to those which were included in the impurity problem. Additionally, we observe orbital selectivity of correlations which further increases the deviation from the renormalized DFT bands.  Still, the spectral function clearly contains sharp bands which we can identify with the DFT bands near the Fermi-level.
    \label{fig:DMFT_spectral_function}
    }
\end{figure*}
{\it Electronic structure at ambient pressure and low temperature.-} At $T_S \approx 90$~K, \ce{FeSe} undergoes a structural phase transition into an orthorhombic phase described by the space group $Cmma$ (No.~67)~\cite{PhysRevLett.114.027001}. The corresponding point group is $D_{2h}$, which, as in the strained case, lacks four-fold rotational symmetry. The electronic band structure of the orthorhombic phase is shown in Fig.~\ref{fig:bandstructures}~c). A zoom of the region along the $\Gamma$–$Z$ path and the corresponding little-group irreducible representations are displayed in Fig.~\ref{fig:zoom}~c). Analogous to strained \ce{FeSe} along the in-plane directions, the reduction in symmetry causes the irreducible representations to subduce to those of a lower-symmetry group, as described in Eq.~\ref{eq:subduce}. Following the same reasoning as in the strained case, we again identify a strong topological insulating phase at the Fermi level.

{\it Symmetry indicators-} We can further refine the topological classification using the irrep decomposition to find the SIs~\cite{mtqc,sis-song}, which can be obtained from the inversion eigenvalues at time-reversal-invariant momenta (TRIMs). For both structures that break the $C_4$ axis, we find at the filling of 28 (Fermi level) $z_{2w,3}=1$ and $z_4=1$ which, although the weak index is odd, indicates a strong TI phase because the $z_4$ index is odd~\cite{mtqc,sis-song}. A strong TI cannot be deformed into a stack of 2D quantum spin Hall insulators and displays surface states on \textit{any} of the surfaces.
Fig.~\ref{fig:surface-states} shows the top surface spectrum for a finite slab perpendicular to ${a}_3$ and along the path the $\frac{2}{5}M-\Gamma-\frac{2}{5}M$ computed with WannierTools for all three structures. For each of them, we identify a 2D Dirac cone at this surface. For the strained tetragonal and the orthorhombic phase, Fig.~\ref{fig:surface-states}~b) and c) respectively, we expect the Dirac cone to be topologically protected as they are strong TIs near the Fermi level. The unstrained tetragonal phase, Fig.~\ref{fig:surface-states}~a), on the other hand does not correspond to a strong TI (as both SIs are zero) and therefore we do not expect the cone to be topologically protected. In Table~\ref{tab:phase_space} we present a summary of the results.

{\it DFT+DMFT calculations.-} To complement our DFT analysis, we perform charge self-consistent DMFT (cscDMFT) calculations in order to examine the effect of electron correlations in the Fe-3$d$-orbitals. Fig.~\ref{fig:DMFT_spectral_function} shows the calculated spectral function of the three different \ce{FeSe} crystal structures considered. As a guideline to the eyes, we visualize the renormalized DFT bands, with average mass-enhancement $Z=2.4$ as indicated in the upper right corner of each subplot, as light blue dashed curves. In line with previous experimental and theoretical results~\cite{yi_observation_2015, Sprau2017} we observe orbital selective correlation effects. The average mass enhancement of $2.4$ induces a shift of the surface Dirac-cone much closer to the experimentally observed value~\cite{edge-states-film}. Furthermore, beyond a simple rescaling of the DFT bands, we observe that the double counting corrections lead to a relative shift between the strongly correlated \ce{Fe}-d orbitals and the weakly correlated \ce{Se} orbitals. This can be most prominently observed along the $\Gamma$-$Z$ line.

Although DFT + DMFT is known to not quantitatively describe the Fermi surface reconstruction in FeSe
~\cite{hubbard-like-bands,emergence-nematic,dichotomy-hole-electron}, the overall spectral redistribution due to correlations has been shown to yield a good qualitative agreement with angle-resolved photon emission spectroscopy (ARPES), giving a general good account of the electronic structure in FeSe~\cite{hubbard-like-bands}. Thus,  on the basis of our cscDMFT results, we expect the topology to be still present, as no new crossings are introduced or the new crossings are known to be irrelevant as in the orthorhombic case, i.e.~the correlation induced crossing brings the orthorhombic case into the same form as the tetragonal structure strained along $a_1$.
Furthermore, the bands near the Fermi level can be identified by the renormalized bands from our DFT calculation, see Fig.~\ref{fig:DMFT_spectral_function}. Admittedly, non-local correlations in \ce{FeSe} are known to change the band structure qualitatively~\cite{nonloc_1, nonloc_2, nonloc_3, nonloc_4, nonloc_5} and we cannot fully rule out the possibility that these missing effect destroys the topology. However, in the $C_4$ broken crystals all bands on all high-symmetry paths have the same symmetry since there is only one possible irrep along the $\Gamma - Z$ line. Therefore, in the process of including renormalization without adding substantial incoherence, the order of the bands is conserved as no crossings occur. Moreover, all the bands involved at the Fermi level have $d_{xz}$, $d_{yz}$ and $d_{x^2-y^2}$ character and are expected to be renormalized in a similar way, as opposed to the $d_{xy}$-band which is more affected by correlations~\cite{yin2011kinetic,hubbard-like-bands,yi_observation_2015}. This suggests that, even though we cannot fully exclude it, the topology is not expected to be destroyed by non-local correlations. 

{\it Summary.-} 
In this work, we have performed \textit{ab initio} DFT calculations for \ce{FeSe} in its tetragonal, orthorhombic, and uniaxially strained tetragonal structures. We classified the resulting electronic structures using TQC and computed their corresponding symmetry indicators. Our analysis reveals that breaking the in-plane four-fold rotational symmetry induces a strong $\mathbb{Z}_2$ topological phase with orthorhombic symmetry at a filling of 28 electrons (Fermi level). To further support this, we identified Dirac-like surface states near the $\Gamma$ point by computing the surface spectral function for a semi-infinite slab geometry along the $a_3$ direction. These calculations were performed using a tight-binding model constructed in a Wannier function basis that accurately reproduces the low-energy band structure of the crystal (see Fig.~\ref{fig:surface-states}). To account for electronic correlation effects in the Fe $3d$ orbitals, we complemented our DFT results with DFT+DMFT calculations. The preservation of coherent low-energy bands and the robustness of the topological classification under symmetry-preserving perturbations that do not close the band gap suggest that the topological character of the system remains intact in the presence of correlations. Taken together, our results establish that \ce{FeSe} lies at the boundary of multiple topological phases and demonstrate that strain is an effective experimental knob to manipulate the system's topological properties.

{\it Acknowledgments.-} This work was supported by the  Deutsche Forschungsgemeinschaft (DFG, German Research Foundation) through QUAST-FOR5249 - 449872909 (project TP4) and through CRC 1487, “Iron,
upgraded!” – project number 443703006. M.G.D.,  and M.G.V. thanks support to PID2022-142008NB-I00 funded by MI-CIU/AEI/10.13039/501100011033 and FEDER, UE, the Canada Excellence Research Chairs Program for Topological Quantum Matter, NSERC Quantum Alliance France-Canada and to Diputación Foral de Gipuzkoa Programa Mujeres y Ciencia. This work was supperted by the IKUR Strategy under the collaboration agreement between Ikerbasque Foundation and DIPC on behalf of the Department of Education of the Basque Government and the Würzburg-Dresden Cluster of Excellence on Complexity and Topology in Quantum Matter, ct.qmat (EXC 2147,
Project ID 390858490) . We acknowledge the support of the Natural Sciences and Engineering Research Council of Canada (NSERC).  This work was supported also by grant 369963 from the Fonds de recherche du Québec.

\appendix
\bibliography{bibliography}

\section{Generating strained structures}
\label{App::relax}
Following the procedure described in the main text, we generate estimates for realistic strained structures by generating strain tensors derived from DFT and applying them to the experimental structure. To do so, we first perform a relaxation at ambient conditions starting from the experimental structure~\cite{topological-nature, Koz2014}. As expected these two structures are \emph{not} identical, see table~\ref{tab:structure}.

\begin{table}[t]
    \centering
    \begin{tabular}{c|c|c|c|c}
       Structure  & a [\AA] & b [\AA] & c [\AA] & $z_{Se}$ \\
       \hline \hline 
       Exp. tet. & 3.7724 & 3.7724 & 5.5217 & 0.2673 \\
       DFT. tet.  & 3.5822 & 3.5822 & 5.1767 & 0.2690 \\ \hline
       Exp. orth. & 5.3104 & 5.3317 & 5.4879 & 0.2662
    \end{tabular}
    \caption{Comparison of the experimental and DFT structure at ambient conditions in the tetragonal phase and the orthorhombic phase all in the respective conventional unit cell.}
    \label{tab:structure}
\end{table}

Starting from this force minimum we found within DFT, we can generate strained structures by fixing certain lattice vectors and relaxing all degrees of freedom but the constrained one. Thereby, we generate a series of structures at different strain conditions, see table~\ref{tab:structure_dft}. After the relaxation we extract the strain tensor and create a strained "experimental" structure.
\begin{table}[t]
    \centering
    \begin{tabular}{c|c|c|c|c|c}
       Strain [\%]  & a [\AA] & b [\AA] & c [\AA] & $z_{Se}$ & $z_{Fe}$ \\
       \hline
       DFT. tet.   & 3.5822 & 3.5822 & 5.1767 & 0.2690 & 0.0000   \\ \hline
       $s_a = -1.0\%$ & 3.5464 & 3.5999 & 5.1900 & 0.2694 & 0.0021 \\
       $s_a = -3.0\%$ & 3.4748 & 3.6349 & 5.2086 & 0.2702 & 0.0061 
    \end{tabular}
    \caption{Relaxed structures from DFT under different stain along the $a_1$-axis in the primitive unit cell.}
    \label{tab:structure_dft}
\end{table}

By comparing the ambient conditions DFT result to the strained structures, we can directly extract the diagonal strain tensor:
\begin{equation}
    \epsilon = \begin{pmatrix}
\epsilon_{xx} & 0 & 0\\
0 & \epsilon_{yy} & 0 \\
0& 0 & \epsilon_{zz}
\end{pmatrix} = \begin{pmatrix}
\frac{a}{a_{\rm{amb.}}} & 0 & 0\\
0 & \frac{b}{b_{\rm{amb.}}} & 0 \\
0& 0 & \frac{c}{c_{\rm{amb.}}}
\end{pmatrix} 
\end{equation}
where $a_{\rm{amb.}},b_{\rm{amb.}},c_{\rm{amb.}}$ are the ambient condition DFT result's lattice parameters and $a,b,c$ are the ones of the strained structure.
From this, we obtain an estimate for an experimental structure under ($L_{\rm{strain}}$) strain by applying the strain tensor we extracted from DFT to the experimental structure at ambient conditions $L_{\rm{Exp.}}$:
\begin{equation}
    L_{\rm{strain}} = \epsilon L_{\rm{Exp.}}
\end{equation}

In other words, we approximate the material's Poisson ratio by the ones predicted by DFT and then use the ratios to estimate structures under strain. A delicate point is how internal coordinates are treated. Optimally, we would be able to relax the internal degrees of freedom after fixing the lattice parameters by the method described above, but the Se position is significantly too small within when relaxing within DFT~\cite{Skornyakov_2018}. Therefore, we again resort to a rescaling. We note that under in-plane strain, the Se position does not drift significantly.

\end{document}